\documentclass[english,twocolumn]{revtex4}
\usepackage[T1]{fontenc}
\usepackage[latin9]{inputenc}
\usepackage{amsmath}
\usepackage{graphicx}
\usepackage{amssymb}

\makeatletter

\usepackage{textcomp}

\makeatletter

\DeclareRobustCommand{\greektext}{%
 \fontencoding{LGR}\selectfont
 \def\encodingdefault{LGR}}
\DeclareRobustCommand{\textgreek}[1]{\leavevmode{\greektext #1}}
\DeclareFontEncoding{LGR}{}{}

\makeatletter



\makeatother

\makeatother

\usepackage{babel}
\makeatother

\begin{document}

\title{Rubidium resonant squeezed light from a diode-pumped optical-parametric
oscillator}

\author{A. Predojevi\'{c}}

\affiliation{ICFO-Institut de Ciencies Fotoniques, Mediterranean Technology Park,
08860 Castelldefels (Barcelona), Spain}

\author{Z. Zhai}

\affiliation{ICFO-Institut de Ciencies Fotoniques, Mediterranean Technology Park,
08860 Castelldefels (Barcelona), Spain}

\author{J. M. Caballero}

\affiliation{ICFO-Institut de Ciencies Fotoniques, Mediterranean Technology Park,
08860 Castelldefels (Barcelona), Spain}

\author{M. W. Mitchell}

\affiliation{ICFO-Institut de Ciencies Fotoniques, Mediterranean Technology Park,
08860 Castelldefels (Barcelona), Spain}

\date{27 July 2008}

\begin{abstract}
We demonstrate a diode-laser-pumped system for generation of quadrature
squeezing and polarization squeezing. Due to their excess phase noise,
diode lasers are challenging to use in phase-sensitive quantum optics
experiments such as quadrature squeezing. The system we present overcomes
the phase noise of the diode laser through a combination of active
stabilization and appropriate delays in the local oscillator beam.
The generated light is resonant to the rubidium D1 transition at 795$\,$nm
and thus can be readily used for quantum memory experiments.
\end{abstract}
\maketitle
Interaction of quantum states of light is of interest both for quantum
communications, for improved sensitivity in measurements limited by
quantum noise, and for understanding light-matter interactions at
the most fundamental level. Our interest is in quadrature squeezing
and polarization squeezing, which are phase-dependent quantum features.
A proven technique for generation of squeezing is phase-sensitive
amplification in a subthreshold optical parametric oscillator (OPO)
\citep{Polzik1992}. For strong interaction with atoms the squeezed
light needs to be atom-resonant, which limits the choice of lasers,
nonlinear crystals, and detectors. Several experiments have demonstrated
squeezing at the rubidium resonance \citep{Akamatsu2004,Appel2008,Hetet2007,Tanimura:06}.
These experiments use distinct methods: squeezing in a waveguide \citep{Akamatsu2004}
and downconversion in an OPO \citep{Tanimura:06,Hetet2007,Appel2008}.
In the latter case, the squeezing at the rubidium D1 line using the
non-linear interaction in a subthreshold OPO was achieved by using
a Ti:Sapphire laser and periodically poled potassium titanium oxide
phosphate (PPKTP) as nonlinear medium \citep{Tanimura:06,Hetet2007,Appel2008}.
The noise suppression of this method was shown to be more than -5dB
\citep{Hetet2007}. To our knowledge the only experiment that generated
squeezing in an OPO pumped by a diode laser system worked at 1080$\,$nm
\citep{Zhang:06}, far from any useful atomic resonance and produced
relative intensity squeezing, a phase-independent property.

Compared to other laser systems, diode lasers are easy to operate,
compact, and inexpensive. It has long been suspected that the excess
phase noise of the diode laser, which results in a relatively large
linewidth, would be an obstacle for production of phase-sensitive
quantum states such as quadrature squeezing. The spectral distribution
of diode laser phase noise over different frequencies was investigated
in \citep{Lax1967}. There is was shown that the main contribution
in the noise comes from the low frequency part of the spectrum, as
expected for a process of phase diffusion. This suggests that the
laser output can be treated as quasi-stationary, with the laser frequency
drifting slowly (on the time-scale of propagation and cavity relaxation)
within the laser linewidth.

Here we show that cavity stabilization of the diode laser frequency,
in combination with appropriate delays for the local oscillator beam,
allows squeezing to be observed with diode laser based systems. The
laser, doubling system, and stabilization use standard techniques
and could be applied to a variety of other wavelengths. The use of
appropriate delays makes the system robust against frequency fluctuations,
and could be incorporated into non-diode-laser pumped systems as well.

We present our experimental system, a PPKTP-based subthreshold OPO
pumped by a frequency-doubled diode laser, and expected and observed
squeezing performance. We then consider the effect of the frequency
fluctuations on the observable squeezing in the regime of the quasi-stationary
fluctuations, an analysis which indicates that the system can be made
immune to random frequency drifts for appropriate local oscillator
delay. Finally, we present measurements of squeezing versus delay
in agreement with the theory.

\section{Experimental setup}

\begin{figure}
\includegraphics{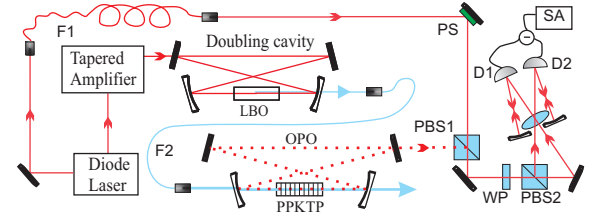}

\caption{Experimental apparatus. Light from the diode laser is amplified in
the tapered amplifier and fed into the doubling cavity. The blue output
light is mode-matched into fiber F2 and fed into the OPO cavity. Both
doubling and OPO cavity are only resonant to the red light. The length
of the local oscillator beam path can be changed by fiber F1. The
modes of squeezed vacuum and local oscillator are then overlapped
on a beamsplitter (PBS1) where power balancing is performed by a waveplate
(WP) and a beamsplitter (PBS2). Light is collected onto diodes D1
and D2 of the balanced detector. The obtained electrical signal is
recorded using a spectrum analyzer (SA).}

\label{fig:experiment}
\end{figure}

The schematic of the experiment is shown in the Fig.\ref{fig:experiment}.
Our laser system (Toptica TA-SHG) consists of a grating stabilized
795$\,$nm diode laser which is amplified by an optical tapered amplifier
and injected into a frequency doubler with lithium triborate crystal
as nonlinear medium. A 20$\,$MHz modulation is applied to the laser
current, resulting in frequency modulation sidebands of 5\%. The reflection
from the cavity is demodulated to provide an error signal (the Pound-Drever-Hall
technique (PDH)). The laser and cavity are locked in frequency by
a proportional-integral-derivative (PID) circuit acting on the cavity
piezo and a fast proportional component acting on the current of the
diode. At the same time, the absolute laser frequency is stabilized
by frequency-modulation spectroscopy of a saturated-absorption signal,
fed back by digital PID to the piezo-electric transducer of the laser
grating. For the experiments described here, the laser was locked
to the $F=2\rightarrow F'=1$ transition of $^{\text{87}}$Rb. Residual
fluctuations of the FM spectroscopy signal indicate that the fast
cavity lock reduces the linewidth to 400$\,$kHz full width half maximum
(FWHM).

The generated 397$\,$nm light is passed through a single-mode fiber
for spatial filtering and pumps the sub-threshold degenerate optical
parametric oscillator. The nonlinear material used in the OPO is a
10$\,$mm long PPKTP crystal, temperature tuned for the maximum second-harmonic
generation efficiency. The OPO cavity is a 64$\,$cm long bow-tie
configuration resonator which consists of two spherical mirrors (R=10$\,$cm)
and two flat mirrors. The distance between the spherical mirrors is
11.6$\,$cm yielding to the beam waist in the crystal of 42$\,$\textgreek{m}m.
The output coupling mirror of the OPO has a transmission of 7.8\%,
and the measured intracavity losses are 0.55\%. The measured cavity
linewidth is $\delta\nu$=8$\,$MHz (FWHM) and the output coupling
efficiency $\eta=0.93$. The free spectral range of the OPO cavity
is 504$\,$MHz. The cavity is locked using the Pound-Drever-Hall technique
performed on the transmission signal of a counter-propagating beam
fed into cavity through the high reflecting flat mirror. The error
signal is digitized and fed into a PID circuit programmed within a
National Instruments field-programmable gate array (FPGA) board type
NI 7833R. It controls the OPO cavity length by moving the position
of one cavity mirror with a piezo-electric transducer.

A local oscillator beam is derived from the diode laser by passing
through single-mode fibers whose combined lengths can be chosen to
give a desired group delay. The vertically-polarized OPO output is
overlapped with 400$\,$\textmu{}W of this horizontally-polarized
beam on polarizing beamsplitter (PBS1). Optimized overlap results
in a measured homodyne efficiency of $\eta_{hom}=0.98$. Local oscillator
and squeezed vacuum beams are mixed and balanced in power on a second
polarizing beamsplitter (PBS2) and detected with a ThorLabs (PDB150)
switchable-gain balanced detector. The quantum efficiency of this
detector at detection wavelength of 795$\,$nm is 88\% by manufacturer
specifications. Losses are mainly caused by the reflection of the
surface of the protective window and diode surface. We use two spherical
mirrors (R=10$\,$mm), to retro-reflect the reflected light onto the
detector improving the quantum efficiency by 7\%, i.e., to 95\%. Quarter-wave
plates are used to prevent the returning light from reaching the OPO
cavity. For the local oscillator power of 400$\,$\textmu{}W electronic
noise of the detector is 14$\,$dB below the standard quantum limit.
Electronic noise was subtracted from all the traces.

\begin{figure}
\includegraphics[scale=0.85]{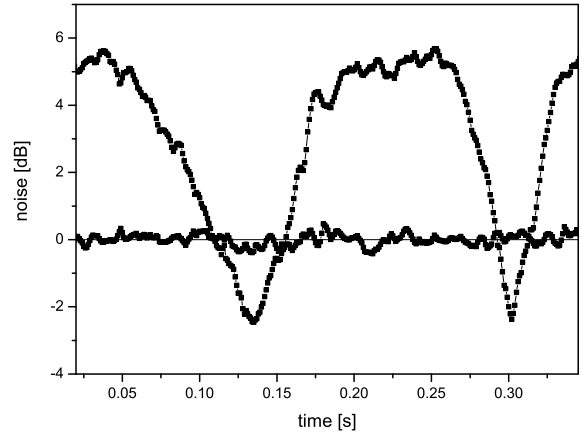}

\caption{Squeezed vacuum generation: (a) squeezing trace when scanning the
phase of the local oscillator (b) shot noise level. Electronic noise
is subtracted. Spectrum analyzer at zero span, resolution bandwidth
= 30kHz, video bandwidth = 30Hz}

\label{fig:sq}
\end{figure}

As described in the theory section, when fluctuations in frequency
are included, the degree of squeezing is expected to depend on the
relative delay through two paths: From laser to PBS1 through the local
oscillator fiber, and from laser to PBS1 through amplifier, doubler,
pump fiber and OPO. Insensitivity to these fluctuations is expected
to occur at a {}``white light'' condition of equal delays. Initial
measurements were taken with the local oscillator fiber chosen to
achieve this condition, as described in detail in section \ref{sec:delay-considerations}.

Noise measurements were performed at fixed frequency, zero-span of
the spectrum analyzer. The degree of squeezing we observe matches
the above mentioned gain and loss parameters for the demodulation
frequencies 3MHz and higher. The highest level of squeezing of -2.5$\,$dB
we observe at the demodulation frequency of 2$\,$MHz shown in Fig.
\ref{fig:sq}.

\section{relative lock quality}

We note that the achieved linewidth for the stabilized diode laser
(400$\,$kHz) is an order of magnitude below the linewidths of the
doubling cavity (14$\,$MHz) and OPO cavity (8$\,$MHz). This justifies
treating the frequency fluctuations of the laser as quasi-stationary
when determining the effect on squeezing. Another treatment of phase
noise has been discussed \citep{Gea-Banacloche1990}, but is far more
involved and does not consider group delay effects. At the same time,
while fast feedback to the laser current allows a high-bandwidth lock
of the laser and doubling cavity, there is no corresponding fast control
for locking of the OPO cavity to the laser frequency. Also, the PDH
scheme, which achieves a very good signal by injecting through the
cavity output coupler, cannot be used in many squeezing experiments
because it would contaminate the squeezed light. For these reasons,
the active stabilization of the OPO cavity may be an important factor
in the performance of squeezing experiments.

\section{Theory}

\begin{figure}
\includegraphics[scale=0.7]{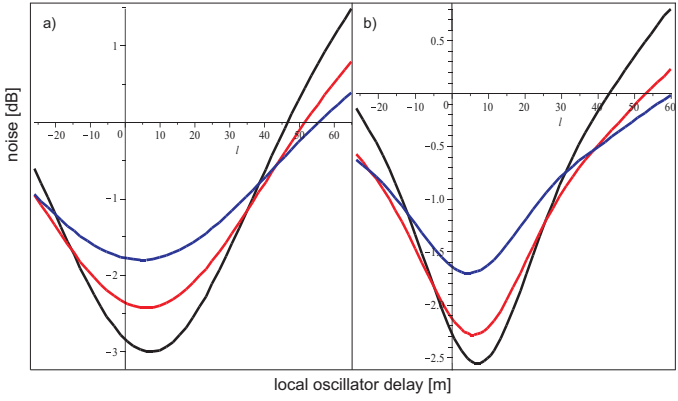}

\caption{Squeezing vs.delay for three different detection frequencies 1$\,$MHz,
2$\,$MHz, and 3$\,$MHz depicted black, red, and blue respectively;
a) models the laser spectrum as Gaussian of linewidth 700$\,$kHz
full width half maximum (FWHM), b) models the laser spectrum as Lorentzian
of linewidth 300$\,$kHz (FWHM).}

\label{fig:gauss}
\end{figure}

The theoretical description adapts the treatment of Collet and Gardiner
\citep{PhysRevA.30.1386} to model the nonlinear interaction inside
the OPO cavity. Here we assume that the frequency drift of the diode
laser is slow on the time scale of the decay of light inside the OPO
cavity. This quasi monochromatic treatment describes a single mode
laser drifting slowly within a finite linewidth. In such a system
frequency fluctuations lead to a fluctuating phase shift between the
squeezed mode and the local oscillator mode. Our calculation modifies
\citep{PhysRevA.30.1386} by including a relative detuning $\Delta\omega$
between pump laser and OPO cavity caused by the random frequency drifts.
As in \citep{PhysRevA.30.1386}, we start from the quantum Langevin
equation of the OPO cavity\begin{equation}
\dot{a}=-\frac{i}{\hbar}[a,H_{sys}]-\left(k_{1}+k_{2}\right)a+\sqrt{2k_{1}}a_{v1}+\sqrt{2k_{2}}a_{v2}\end{equation}
 where $a$ and $a^{\dagger}$ denote annihilation and creation operators
of the cavity mode with frequency $\omega_{0}$, $k_{1}$ and $k_{2}$
denote the loss rates due to output coupler and intracavity losses,
and $a_{v1}$ and $a_{v2}$ denote the annihilation operators of the
(vacuum) field entering the cavity due to output coupler and intracavity
losses. The Hamiltonian operator of the system is\begin{equation}
H_{sys}=\hbar\omega_{0}a^{\dagger}a+\frac{i\hbar}{2}(\epsilon e^{-i\omega_{P}t}(a^{\dagger})^{2}-\epsilon^{*}e^{i\omega_{P}t}a^{2})\end{equation}
 where the first term describes the energy of photons inside the cavity
while the second term models the non-linear interaction induced by
the pump field with frequency $\omega_{P}$. The phase of the non-linear
coupling constant $\epsilon=\left|\epsilon\right|e^{i\phi}$ is determined
by the phase of the pump field, $\phi$. Furthermore, we assume that
the squeezed mode is detuned from the cavity resonance by $2\Delta\omega\equiv\omega_{P}-2\omega_{0}$.
By performing the equivalent calculation as in \citep{PhysRevA.30.1386}
we finally reach the Bogoliubov transformation from input to output
fields

\begin{eqnarray}
\tilde{a}_{out}\left(\omega+\Delta\omega\right) & = & \left[A_{1}\tilde{a}_{v1}\left(\omega+\Delta\omega\right)+A_{2}\tilde{a}_{v2}\left(\omega+\Delta\omega\right)\right.\nonumber \\
 &  & +\left.C_{1}\tilde{a}_{v1}^{\dagger}\left(-\omega+\Delta\omega\right)\right.\nonumber \\
 &  & +\left.C_{2}\tilde{a}_{v2}^{\dagger}\left(-\omega+\Delta\omega\right)\right]B^{-1}\label{eq:outputfield}\end{eqnarray}

where \begin{eqnarray}
A_{1} & = & \eta^{2}-\left(1-\eta-i\Omega\right)^{2}+\Delta\Omega\left(2\eta i-\Delta\Omega\right)+\left|\alpha\right|^{2}\\
A_{2} & = & 2\sqrt{\eta\left(1-\eta\right)}\left[i\left(-\Omega+\Delta\Omega\right)+1\right]\\
C_{1} & = & 2\eta\alpha\\
C_{2} & = & 2\alpha\sqrt{\eta\left(1-\eta\right)}\\
B & = & \left(1-i\Omega\right)^{2}+\Delta\Omega^{2}-\left|\alpha\right|^{2}\end{eqnarray}

We have introduced $\tilde{a}$ as the operators in rotating frame,
and scaled all frequencies and rates to the cavity linewidth, i.e.
demodulation (detection) frequency $\Omega=\frac{\omega}{k_{1}+k_{2}}$,
detuning $\Delta\Omega=\frac{\Delta\omega}{k_{1}+k_{2}}$, cavity
escape efficiency $\eta=\frac{k_{1}}{k_{1}+k_{2}}$, $1-\eta=\frac{k_{2}}{k_{1}+k_{2}}$,
and pump amplitude $\alpha=\frac{\varepsilon}{k_{1}+k_{2}}$. The
squeezing spectrum $S(\Omega)$ can be deduced from equation (\ref{eq:outputfield})
using $q_{\theta}=\frac{1}{\sqrt{2}}(\tilde{a}_{out}e^{-i\theta}+\tilde{a}_{out}^{\dagger}e^{i\theta})$

\begin{eqnarray}
S(\Omega) & = & 1+2\eta_{det}\left\langle :q_{\theta},q_{\theta}:\right\rangle \nonumber \\
 & = & 1+\frac{8\eta_{det}\eta\left|\alpha\right|^{2}}{\left|B\right|^{2}}\left[1+\frac{B}{2|\alpha|}\cos\left(\Delta\phi+2\Delta\omega\tau_{D}\right)\right.\nonumber \\
 &  & +\left.\frac{\Delta\Omega}{|\alpha|}\sin\left(\Delta\phi+2\Delta\omega\tau_{D}\right)\right]\end{eqnarray}

where $\theta=\theta_{0}+\Delta\omega\tau_{D}$ denotes the phase
of the local oscillator, with $\theta_{0}$ being the phase of the
local oscillator in the white light configuration and $2\Delta\omega\tau_{D}$
being the phase shift for detuned local oscillator when the light
is delayed for $\tau_{D}$ from the white light configuration. Furthermore,
$\Delta\phi=2\theta_{0}-\phi$ denotes the relative phase between
the phase of the local oscillator in the white light configuration
and of the pump laser of the OPO, $\alpha=|\alpha|e^{i\phi}$. Best
squeezing is obtained for the phase that gives

\begin{equation}
\tan(\Delta\phi+2\Delta\omega\tau_{D})=\frac{2\Delta\Omega}{1-\Delta\Omega^{2}+\Omega^{2}+|\alpha|^{2}}\end{equation}

which due to the cavity dispersion depends on the detuning of the
pump laser $\Delta\Omega$. The right side of the equation represents
the delay in the OPO cavity. In first order of the detuning the squeezing
phase is

\begin{equation}
\Delta\phi+2\Delta\omega\tau_{D}=\pi+\frac{2}{1+\Omega^{2}+|\alpha|^{2}}\Delta\Omega\end{equation}

This dispersion can be compensated by delaying the local oscillator
before the homodyne detection. A delay line of length $l$ and group
index $n_{g}$ will introduce the phase shift

\begin{equation}
2\Delta\omega\tau_{D}=2(k_{1}+k_{2})\frac{ln_{g}}{c}\Delta\Omega\end{equation}

Thus for delay length

\begin{equation}
l=\frac{c}{n_{g}(k_{1}+k_{2})(1+\Omega^{2}+|\alpha|^{2})}\label{eq:delaylength}\end{equation}

the homodyne detection will be performed, to first order, at the correct
squeezing phase $\Delta\phi=\pi$ even for detuned pump. The dispersion
in the OPO cavity and therefore also the compensation length depends
on the detection frequency $\Omega$. For higher detection frequency
a shorter compensation delay is necessary. Assuming a slowly drifting
laser with power spectral density of $\rho(\Delta\Omega)d\Delta\Omega$
the obtained squeezing can be modeled by averaging the homodyne power
spectrum $S(\Omega)$ for phase $\Delta\phi=\pi$ over $\Delta\Omega$.
The averaged squeezing spectrum

\begin{equation}
\bar{S}(\Omega)=\int_{-\infty}^{+\infty}S(\Omega,\Delta\Omega)\rho(\Delta\Omega)\, d\Delta\Omega\end{equation}

is plotted in Fig.\ref{fig:gauss} for a Gaussian and a Lorentzian
linewidth $\rho(\Delta\Omega)$. We note that physically $\Delta\Omega$
is the mismatch between half the pump frequency and the OPO cavity
frequency scaled to the cavity linewidth, and thus both laser frequency
fluctuations and OPO cavity fluctuations will contribute to $\rho(\Delta\Omega)$.
The shift of optimum squeezing to positive delay is due to the existence
of the delay introduced by OPO.

\section{delay considerations\label{sec:delay-considerations}}

We note that in Eqs. (\ref{eq:outputfield}) and (\ref{eq:delaylength}),
$\tau_{D}$ is the group delay between the local oscillator and the
pump light at the cavity. As both local oscillator and pump are ultimately
derived from the same laser, we can identify $\tau_{D}=0$ as a {}``white-light''
condition in a Mach-Zehnder-topology interferometer. The light in
the squeezing path passes the tapered amplifier, doubling cavity,
mode matching fiber, lengths of free-space propagation, and the OPO
cavity. The light in the local oscillator path passes lengths of free-space
propagation and a mode-cleaning fiber (which we use to introduce the
desired delays).

In presenting the results, the {}``zero'' of $\tau_{D}$ is taken
to be when the total delay in the local oscillator path, as calculated
from measurements of fiber and free-space lengths, is equal to the
combined delays in the amplifier, doubling cavity, fiber and free
space. We do not include the OPO cavity delay because this depends
on $\Omega$, as presented in the theory and shown in fig. \ref{fig:gauss}.

The delay introduced by the doubling cavity is the cavity group delay
at line center \begin{equation}
\tau=\frac{1}{\pi\cdot\delta\nu}\end{equation}
 The measured doubling cavity linewidth is $\delta\nu=14$$\,$MHz.
To delay the local oscillator we have used fibers with group index
$n_{g}=1.5$.

We note that, as the the laser and doubling cavity are mutually locked,
it is not obvious how the doubling cavity delay should be included.
While the light is obviously propagating from laser through doubling
cavity, a frequency fluctuation in the doubling cavity will, via the
current feedback, affect the laser frequency. We choose to include
the cavity delay, in the squeezing path because it gives best agreement
with the data presented below.

\section{measured squeezing vs. delay}

\begin{figure}
\includegraphics{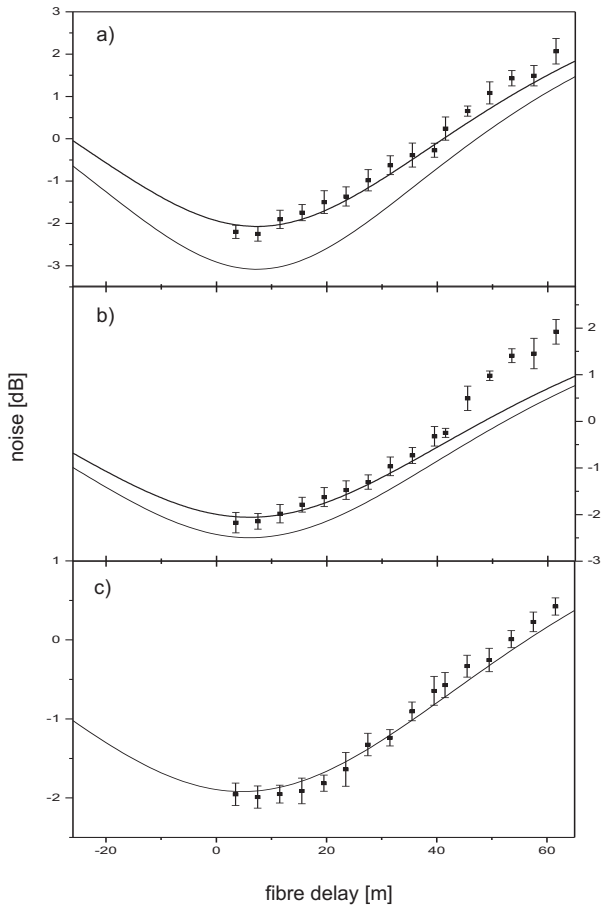}

\caption{Squeezing dependent on the path mismatch measured for three
different detection frequencies: a) at 1$\,$MHz demodulation frequency,
b) at 2$\,$MHz demodulation frequency c) at 3$\,$MHz demodulation
frequency. The points show the experimental data, the solid lines
the predicted level of squeezing for the parameters measured in the
experiments using a Gaussian profile of 700$\,$kHz linewidth (FWHM),
the dashed lines show the theoretical level of squeezing for the same
parameters as the solid line with additional technical noise independent
of the relative delay. The error bars represent standard deviation
over series of identical measurements.}

\label{fig:delay}
\end{figure}

We have performed a series of measurement where a controllable delay
was introduced in the path of the local oscillator with intention
to: (i) measure level of squeezing in white light configuration (ii)
see the effect of the change of delay on the level of squeezing. The
results are presented in the Fig. \ref{fig:delay}.

We performed the measurements of the quadrature variance for every
four meters added in the local oscillator path starting from the proximity
of the balanced delay configuration. Final fiber length was 60$\,$m
longer than the balanced configuration. Due to the limited pump power
and large fiber losses for the blue light measurements at negative
delay were not feasible. We measured squeezing vs. delay for three
different demodulation frequencies 1$\,$MHz, 2$\,$MHz and 3$\,$MHz
(Fig. \ref{fig:delay}).

The experimental results show minima at positive delay as predicted
by theory. Eqn. \ref{eq:delaylength} predicts $l=\{7.3,6.0,4.7\}$$\,$m
shift for demodulation frequencies $\{1,2,3\}$$\,$MHz, respectively.
Here we assume that the doubling cavity delay is equal to the delay
which the cavity introduces at the resonance. Naturally this delay
does not depend on the demodulation frequency.

The theoretical curves in Fig. \ref{fig:delay} are obtained using
all experimental parameters as stated above, but varying the width
of $\rho(\Delta\Omega)$ as the only free parameter. Of two different
profiles treated in the theory the comparison with the Gaussian reflects
the shape of the experimental curve more closely than the Lorentzian
profile. We see good agreement, especially at 3$\,$MHz demodulation
frequency, for a Gaussian spectrum of 700$\,$kHz (FWHM). Using the
in-loop signal from the laser lock to a saturated-absorption reference,
we find a 400$\,$kHz laser linewidth. A similar measurement of the
distribution of $\omega_{0}-\omega_{laser}$ can be made using the
OPO cavity locking signal. Under the conditions of the squeezing measurements,
however, the locking signal was too weak to extract a meaningful signal,
largely because we cannot inject through the output mirror as in the
PDH technique. We can place a lower limit of 300$\,$kHz on the width
of $\rho(\Delta\Omega)$ based on PDH locking of the same cavity,
and the 700$\,$kHz estimate for the width of $\rho(\Delta\Omega)$
appears reasonable.

On the other hand, the level of squeezing we observe in the 1$\,$MHz
and 2$\,$MHz measurements is smaller than predicted by theory. This
might be caused by the light back reflected form the n-faces of the
nonlinear crystal contaminating the squeezed light. If we assume that
this noise is independent of the relative delay, it can be modeled
by a constant offset to our theoretical squeezing curves. With an
offset of (+0.07,+0.03) relative to the standard quantum limit for
(1, 2) MHz, respectively, the theory for a Gaussian laser spectrum
of 700$\,$kHz fits well in shape and amplitude to our measured data
as shown in fig. \ref{fig:delay}. By solving the problem of noise
which causes the decrease of squeezing in the 1$\,$MHz and 2$\,$MHz
measurements one could in agreement with the theory detect more than
5$\,$dB of noise reduction at the OPO output.

\section{conclusion}

We have demonstrated quadrature and polarization squeezing using a
sub-threshold OPO and a frequency-doubled diode laser for a pump.
We have investigated and optimized the squeezing properties by using
a delayed local oscillator. We adapted the theoretical description
of Collet and Gardiner \citep{PhysRevA.30.1386} under the assumption
of slow frequency fluctuations. The theoretical description can be
used to model random frequency fluctuations of the laser but also
the problem of optimization of the OPO cavity stabilization. This
approach showed that the OPO cavity exhibits dispersive behavior which
causes a delay of the squeezed light. Optimum squeezing is observed
if the squeezed light is in white light configuration with respect
to the local oscillator. Experimental results confirmed that the vacuum
mode of the OPO is also taking part in the delay line. This investigation
shows that, by taking into account the balancing and the delay lines,
diode laser sources can be used for producing quadrature and polarization
squeezing in an OPO. Since diode lasers are much cheaper and simpler
to operate our work brings portable inexpensive squeezing devices
for application in e.g.$\,$ precisions measurements into reach.

We gratefully acknowledge inspiring discussions and motivating support
by Eugene S. Polzik , Professor (NBI Copenhagen).

This investigation was supported by the Departament d'Universitats,
Recerca i Societat de la Informació of the Generalitat de Catalunya,
the European Social Fund and the Ministerio de Educación y Ciencia
under the FLUCMEM project (Ref. FIS2005-03394), the Consolider-Ingenio
2010 Project {}``QOIT'' and by Marie Curie RTN {}``EMALI.''

\bibliographystyle{plainnat}

\end{document}